\documentclass[prb,twocolumn,superscriptaddress,showpacs,floatfix,longbibliography]{revtex4-2}
\usepackage{mathrsfs,braket}
\usepackage{amssymb, amsbsy, amsmath,graphicx,mathrsfs,color,ulem,bm}
\usepackage{multirow}
\usepackage{comment,xcolor}
\usepackage{physics}
\usepackage[colorlinks=true,citecolor=blue,urlcolor=blue]{hyperref}

\newcommand{\n}{ \notag \\ }
\begin{document}
	\title{ Luther-Emery liquid in one-dimensional attractive Fermi gas  }
	\author{Hai-Ying Cui}
	\affiliation{Innovation Academy for Precision Measurement Science and Technology, Chinese Academy of Sciences, Wuhan 430071, China}
	\affiliation{University of Chinese Academy of Sciences, Beijing 100049, China}
	
	\author{Yi-Cong Yu}
	
	\affiliation{Innovation Academy for Precision Measurement Science and Technology, Chinese Academy of Sciences, Wuhan 430071, China}

	\author{Aashish Kafle}
	\affiliation{Department of Physics and Astronomy,  Rice University, Houston, Texas 77251-1892, USA}

	\author{Han Pu}
\affiliation{Department of Physics and Astronomy,  Rice University, Houston, Texas 77251-1892, USA}

	\author{Randall G. Hulet}
\affiliation{Department of Physics and Astronomy,  Rice University, Houston, Texas 77251-1892, USA}

\author{Thierry Giamarchi}
\email[]{Thierry.Giamarchi@unige.ch}
\affiliation{Department of Quantum Matter Physics, University of Geneva, Geneva, Switzerland}

\author{Xi-Wen Guan}
\email[]{xiwen.guan@anu.edu.au}
\affiliation{Innovation Academy for Precision Measurement Science and Technology, Chinese Academy of Sciences, Wuhan 430071, China}

\affiliation{Hefei National Laboratory, Hefei 230088, People’s Republic of China}

\affiliation{Department of Fundamental and Theoretical Physics, Research School of Physics, Australian National University, Canberra, ACT 0200, Australia}

\begin{abstract}
One-dimensional (1D) interacting Fermi gases underpin a rich variety of quantum liquids, ranging from the Tomonaga-Luttinger liquid and the Fulde-Ferrell-Larkin-Ovchinnikov paired state to the Luther-Emery liquid (LEL). 
Notably, the LEL-characterized by gapless excitations in charge or spin  degree of freedom  and gapped excitations in the other-has long been theoretically predicted but has eluded definitive experimental observation to date. 
By exactly solving the attractive Yang-Gaudin model via the Bethe ansatz, we fully determine the low-energy excitation spectra for both spin and charge channels. 
For fully paired fermions, the spectrum features gapped spin excitation coexisting with gapless charge excitation. 
We further employ bosonization to derive analytical expressions for the dynamical structure factor (DSF) of the gapped spin excitation, which is a long-standing challenge  in this study. 
In the trapped gas, we find that the spin Bragg spectroscopy peak occurs at a much higher frequency than the charge-mode peak, where  the splitting frequency between these two peaks scales approximately linearly with the interaction strength.
Crucially, the spin gap manifests as a sharp threshold in the spin DSF, below which spectral weight is completely absent, proving a direct and unambiguous spectroscopic fingerprint of the LEL. 
Our work establishes quasi-1D trapped  attractive Fermi gases as a pristine experimental platform for the long-sought direct observation of the LEL.

	\end{abstract}
	\maketitle
	\section{Introduction}
One-dimensional (1D) interacting many-body systems exhibit rich universal collective behavior, starting with the pioneering studies by Tomonaga and Luttinger~\cite{Tomonaga:1950,Luttinger:1963}, and have yielded fundamental insights into the physics of strong correlations. 
Driven by sustained theoretical advances-including perturbative renormalization group theory~\cite{Solyom:1979}, Bethe ansatz~\cite{Guan:2013}, and bosonization~\cite{Haldane:1980,Haldane:1981,Giamarchi:2004,Giamarchi:1988,Schulz:1998,Gogolin:1999,Nagaosa:1999,Schoeller:1998,Tognetti:2000,Cazalilla:2011,Lebrat:2018}-a vast family  of distinct 1D quantum liquids  has been identified.

In 1D repulsive fermionic systems, such as the Yang-Gaudin and Hubbard models, low-energy excitations separate into two fractionalized collective modes carrying solely spin or charge quantum numbers, an iconic hallmark of spin-charge separation unique to the Tomonaga-Luttinger liquid (TLL)~\cite{Tomonaga:1950,Luttinger:1963}.
The TLL~\cite{Haldane:1980,Haldane:1981,Giamarchi:2004} serves as a paradigmatic quantum critical phase, featuring gapless spin and charge excitations that propagate at different velocities.
When backscattering flows to strong coupling, a gap emerges in the spin channel (Umklapp scattering can instead open a charge gap in the 1D Hubbard model)~\cite{Giamarchi:2004}.
Luther and Emery showed that the Hamiltonian governing this gapped sector reduces to a bilinear expression of spinless fermions, which is exactly diagonalizable at the free-fermion Luttinger parameter $K=1/2$ ~\cite{Luther_Emery:1974}.
This phase-hosting one gapless and one gapped excitation branch-is known as the Luther-Emery liquid (LEL).
The LEL has been predicted across numerous correlated many-body platforms: the half-filled Hubbard model~\cite{G_XianLong:2007}, hole-doped Haldane spin-$1$ chains~\cite{Haldane_spin_1:2024}, carbon nanotubes~\cite{Kuzmany:2007}, and trapped 1D Fermi gases~\cite{Fuchs:2004}.
The emergent excitation gap breaks the canonical TLL description: bosonic density fluctuations, massless within pure TLLs, acquire a finite mass. 
Correlation functions no longer follow power-law decay and instead decay exponentially, substantially complicating analytical treatments.
Furthermore, the LEL exhibits a suite of unconventional emergent phenomena.
Reference~\cite{Ying-Ming Xie:2025} classifies the LEL as a quasi-one-dimensional supersolid. 
Its antiparallel-spin fermion pairing bears close analogy to pairing in conventional superconductors, marking the LEL as a one-dimensional analog of three-dimensional superconducting states~\cite{G_XianLong:2007,Orignac:2003}.
Nevertheless, despite decades of exhaustive theoretical work, direct experimental detection of the Luther-Emery phase remains elusive to date.

	 
Building on  ultracold atom manipulation,  Bose and Fermi gases can now be confined in artificial crystal lattices, providing nearly ideal and highly tunable platforms for strongly interacting quantum systems. 
Among these, the 1D Fermi gas stands out as a canonical platform that bridges theory and experiment~\cite{Guan:2013,Lebrat:2018}.  
Recently, the 1D repulsive Fermi gas has been used to investigate the spin-charge separation~\cite{Giamarchi:2004,Hilker:2017,Salomon:2019,Hulet:2018,Hulet:2022,Vijayan:2020}. 
Especially in Ref. \cite{Hulet:2022}, one used Bragg spectroscopy to measure the dynamic structure factors (DSF) of the low-energy spin  density waves (SDW) and charge density wave(CDW) modes. 
The precise control and quantitative characterization of system parameters in such experiments not only provide new physical insights but also remarkably  illustrate the power of quantum simulation with ultracold atoms. 
Bragg spectroscopy has thus  served  as a powerful tool for probing low-energy excitations in strongly correlated many-body systems.
 In the presence of attractive interaction, the 1D Fermi gas  is predicted to form a peculiar 1D liquid phase characterized by a massive spin sector, i.e., a LEL \cite{Luther_Emery:1974}. 
 Furthermore, Ref.~\cite{Fuchs:2004,Tokatly:2004,Batchelor-Guan:2026} demonstrates that an integrable model of two-component attractive Fermi gases in a quasi-1D geometry exhibits 
 a 1D version of the BEC-BCS crossover.

 In this letter, by means of Bethe ansatz exact solution, 
  we rigorously  determine the spin and charge  low-energy excitation spectra for fully paired fermions, showing  gapped spin excitations coexisting with gapless charge excitations.
The spin and charge dispersions reveal significant feature of the LEL by providing energy gap, spin and charge velocities. 
The central theoretical challenge lies in computing the DSF for gapped excitations, a long-standing problem for which only limited analytical results exist~\cite{,Kuzmany:2007,Orgad:2001,Egger:1998,Voit:1996,Essler:2002,Mori:1996}. 
To this end, we  derive an analytical expression for the gapped spin DSF according to the Luther-Emery solution. 
Combined with the exact low-energy excitation spectra obtained from the Bethe Ansatz, our finding  enables  us to independently measure the DSF of the spin and charge modes, which are predicted to exhibit strikingly different spectral features-thus providing a clear and unambiguous experimental signature of the LEL. 
 Finally, we  propose an experimental  accessible scheme to observe the LEL in a trapped 1D two-component attractive Fermi gas using Bragg spectroscopy.


	\section{Yang-Gaudin model and Luther-Emery phase}\label{Section II}
	\subsection{Model}
 We consider a 1D delta-function interacting spin 1/2 Fermi gas in an attractive regime, the Hamiltonian is given by
\begin{eqnarray}
	H=-\frac{\hbar^2}{2m}\sum_{i=1}^{N}\frac{\partial^{2}}{\partial x_{i}^{2}}+2c\sum_{i,j=1}^{N}\delta(x_i-x_j)-hM-\mu N, \label{Hamiltonian}
\end{eqnarray} 
where $N=N_\uparrow+N_\downarrow$  is the total particle number and $M=N_\uparrow-N_\downarrow$ is the magnetization, with $N_\sigma$ ($\sigma = \uparrow,\,\downarrow$) the number of fermions in each spin state. The parameters $h$ and $\mu$ denote the external effective Zeeman ield and chemical potential, respectively. This model, known as the Yang–Gaudin model, was exactly solved by Yang \cite{C.N.Yang:1967} and Gaudin  \cite{Gaudin:1967} using the Bethe Ansatz. For convenience, we set $\hbar=2m=1$ throughout this work, and we will focus on attractive interaction with $c<0$. The coupling strength is described by the quantity $\gamma\equiv c/n$ where $n$ is the density of the particle number.
The universal low-energy properties of the system are captured by the thermodynamic Bethe Ansatz (TBA) equations. In the zero-temperature limit, the TBA equations reduce to a set of coupled integral equations for the dressed energies \cite{Takahashi:1994,Lai:1973,Lai:1971,Guan:2013}.
	\begin{eqnarray}\label{TBA_eq}
	\begin{split}
		\epsilon_b(\Lambda)&=2\left(\Lambda^2-\mu-\frac{c^2}{4}\right)+\int^{B}_{-B} a_2(\Lambda-\Lambda^{\prime}) \epsilon_b(\Lambda^{\prime})d\Lambda^{\prime}\\
		&\ -\int^{Q}_{-Q}a_1(\Lambda-k)\epsilon_u(k)dk,\\
		\epsilon_u(k)&=\left(k^2-\mu-\frac{h}{2}\right)-\int^B_{-B}a_1(k-\Lambda)\epsilon_b(\Lambda)d\Lambda.
	\end{split}
\end{eqnarray}
  Here $\epsilon_{u(b)}$ are the dressed energies of the unpaired(paired) fermions, respectively, with $k,\Lambda$ the corresponding rapidities. The integration boundaries $B$ and $Q$ define the Fermi surfaces for bound pairs and unpaired fermions. The kernel functions are given by $a_n(x)=(1/2\pi)\left\{(nc)/\left[(nc)^2/4+x^2\right]\right\}$. The dressed energies satisfy $\epsilon_u(k)\leqq0\left[\epsilon_b(\Lambda)\leqq0\right]$ for $|k|\leqq Q(|\Lambda|\leqq B)$ corresponds to the occupied states; the positive parts describe unoccupied states.
  The rapidity distributions $\rho(k)$ and $\sigma(\Lambda)$ for unpaired and paired fermions are determined by the Bethe Ansatz integral equations in the thermodynamic limit
  \begin{eqnarray}\label{BA_eq}
  	\begin{split}
  		\rho(k) & =\frac{1}{2 \pi}-\frac{1}{2 \pi} \int_{-B}^B a_1(k-\Lambda)\sigma(\Lambda) d \Lambda, \\
  		\sigma(\Lambda) & =\frac{1}{\pi}-\frac{1}{2 \pi} \int_{-B}^B a_2(\Lambda-\Lambda^{\prime})\sigma(\Lambda^{\prime})d \Lambda^{\prime}\\
  		\ \ &-\frac{1}{2 \pi} \int_{-Q}^Q a_1(\Lambda-k)\rho(k) d k.
  	\end{split}
  \end{eqnarray}
  The densities of unpaired and paired fermions are then obtained as
  \begin{eqnarray}\label{n_b_b_u}
  	\begin{split}
  		n_u=\ \int^{Q}_{-Q}\rho(k)dk,\ \ n_b=2\int^B_{-B}\sigma(\Lambda)d\Lambda.
  	\end{split}
  \end{eqnarray} 
  Solving the Eqs.\eqref{TBA_eq} and \eqref{BA_eq} yields the zero-temperature phase diagram of the Yang-Gaudin model in the $\mu-h$ plane \cite{Guan:2011}, which comprises three distinct phases: a fully paired phase, a fully polarized phase, and a partially polarized phase. 
  The low-energy excitations and physical properties differ dramatically among these phases.
   In this work, we focus exclusively on the fully paired phase, where all fermions form two-body bound states with antiparallel spins, i.e.,  the number of unpaired fermions $N_u=0$ and the system is spin-balanced.  As we shall demonstrate below, the low-energy excitations in this phase exhibit the characteristic signature of a LEL: gapped spin excitations coexisting with gapless charge excitations \cite{Giamarchi:2004,Fuchs:2004}.
  
  The discussion above pertains to a homogeneous system. In the experiment, however, the fermions are confined in a quasi-1D harmonic tube by an inhomogeneous potential. 
  To account for the trapping potential, we employ the local density approximation (LDA), which treats the system locally as a homogeneous gas with a spatially varying chemical potential $\mu(x)=\mu_0-V(x)$, where $\mu_0$ is the chemical potential at the center and $V(x)=1/2m\omega_x^2x^2$ is the axial trapping potential.
   This approach allows us to connect the exact TBA results for the uniform system to the experimentally relevant trapped configuration.
  
  \subsection{Low-energy excitations and the Luther-Emery liquid}
\begin{figure}[!]
   	\centering
   	\includegraphics[scale=0.61]{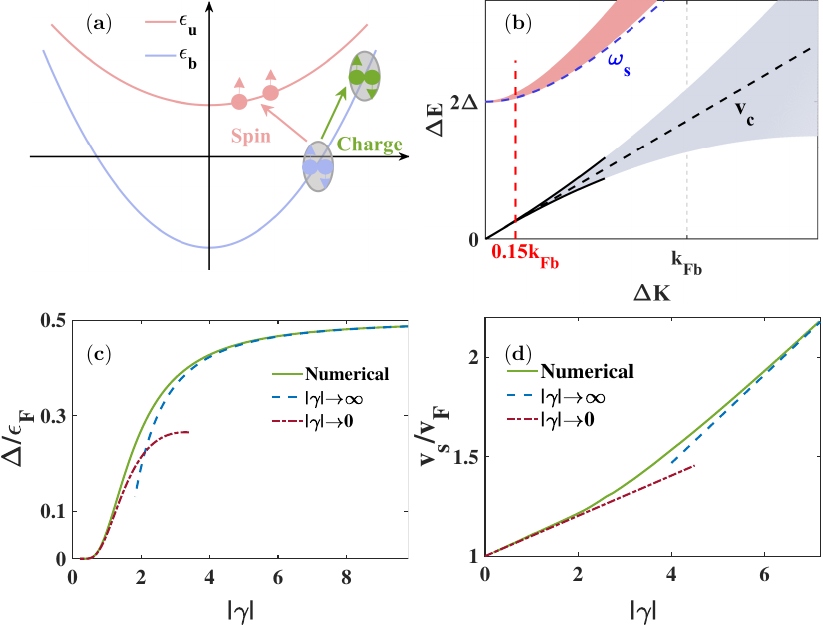}
   	\caption{(a) Schematic illustration of spin and charge excitations in the fully paired phase. The blue and pink curves denote the dressed energies of paired and unpaired fermions, respectively. The gray oval represents a bound state formed by two antiparallel-spin fermions and the arrows indicate their spin orientations; the bule spheres represent the paired fermions within the Fermi point. Charge excitations correspond to particle-hole processes that shift bound states outward from the Fermi points without breaking the pairs (green spheres). 
In contrast, spin excitations break a bound state and flip the spin orientation (pink spheres).  (b) Exact low energy excitation spectra in charge (black) and spin (pink) sectors at $\gamma= c/n=-2.62$ with the Fermi point $k_{Fb}=\pi n_b$. 
The grey area shows the  particle-hole gapless excitation spectrum in long wave  length limit.
 The black solid lines indicate the thresholds of particle-hole excitation which remarkably manifest the free fermion-like dispersion Eq.\eqref{Charge_Dispersion} with an effective mass $m^*\simeq1.46m$ for $\gamma=-2.62$ and the momentum $k=0.15k_{Fb}$.
  The black dashed line in the charge excitation spectrum represents the charge velocity $v_c$. 
  The pink area corresponds to the spin excitations with pair-breaking, and the blue dashed line stands for the gapped dispersion Eq.~\eqref{spin_dispersion}. 
  The red dashed line indicates the positions of excitation momenta in charge and spin sectors for Fig.~\ref{Fig2}(b) and (d). (c) and (d) are the comparison between our numerical results (green solid line) and the previous Bethe ansatz  results\cite{Fuchs:2004}. 
  $v_F=\pi n$ and $\epsilon_F=\pi^2n^2/4$ are respectively the Fermi velocity  and Fermi energy. 
  The blue dashed lines and red dotted lines represent the theoretical result in the limit of $|\gamma|\rightarrow\infty$ and $|\gamma|\rightarrow0$, respectively.}
   	\label{Fig1}
   \end{figure}

   We now turn to the low-energy excitations in the fully paired phase. By solving the Bethe Ansatz equations \eqref{TBA_eq} and \eqref{BA_eq}, we obtain the exact excitation spectra and dispersion relations for both spin and charge sectors. 
    The excitations in these two sectors exhibit fundamentally different behaviors, as illustrated schematically in Fig.~\ref{Fig1}(a). 
    In the charge sector, excitations correspond to particle-hole processes that shift bound pairs outward from the Fermi points without breaking them. In stark contrast, spin excitations necessarily involve breaking a bound pair and promoting the two fermions to unpaired levels, which opens a gap in the spin spectrum. 
   The numerically exact low-lying excitation spectra are shown in Fig.~\ref{Fig1}(b).  For the charge sector, the excitations remain gapless, with the low-energy dispersion precisely given by
   \begin{eqnarray}\label{Charge_Dispersion}
   	\omega_c(q)=v_c|q|\pm\frac{q^2}{2m^{*}}+\cdots,
   \end{eqnarray}
   where the sound velocity $v_c$ and effective mass $m^{*}$ are respectively determined by
   \begin{eqnarray}
   	v_c&=&\frac{\epsilon_b^{\prime}(B)}{2\pi\sigma(B)},\\
   	m^{*}&=&\left[\frac{\epsilon_b^{\prime\prime}(B)}{2\pi^2\left[\sigma(B)\right]^2}-\frac{\sigma^{\prime}(B)\epsilon_b^{\prime}(B)}{2\pi^2\left[\sigma(B)\right]^3}\right]^{-1}.
   \end{eqnarray}
   The derivation of $v_c$ and $m^{*}$ follows the procedure outlined in \cite{J_F_Pan:2023}.
For small momentum $q$,   charge excitations exhibit a linear dispersion relation $\omega_c=v_c|q|$.
This behavior is fully reproduced by the leading-order term in Eq.\eqref{Charge_Dispersion}. 
 Whereas the second-order contribution is negligible.
We emphasize that the charge-sector spectra of the one-dimensional attractive Fermi gas with finite polarization \cite{J_F_Pan:2023} and the unpolarized counterpart share identical gapless characteristics.
 Nevertheless, we will demonstrate that their spin-sector spectra differ substantially, even at infinitesimally small polarization.
This reflects the intrinsic nature of collective motion that arises uniquely in one dimension.

      For the spin sector, the excitation spectrum is gapped and  the dispersion relation takes the relativistic form 
   \begin{eqnarray}\label{spin_dispersion}
   \omega_s(q)=\sqrt{4\Delta^2+\left(v_sq\right)^2}
   \end{eqnarray}
   in the limit of $q\rightarrow0$, as indicated by the blue dashed line in Fig.~\ref{Fig1}(b). 
   The sound velocity $v_s$ and the energy gap $\Delta$ have been previously obtained analytically in the weak-and strong-coupling limits~\cite{Fuchs:2004}. 
   Here, we extend these results by numerically computing $\Delta$ and $v_s$ over the entire range of coupling strength $\gamma$ by solving  Eq.~\eqref{TBA_eq}.  Our results are compared with the previous analytical predictions in Fig.~\ref{Fig1}(c) and (d), showing excellent agreement in the appropriate limits.
   
      A key observation is that the spin gap increases monotonically with $\gamma$ but becomes exponentially small when $\gamma$ is small. 
   This makes it extremely difficult to detect the gapped nature of the LEL in the weak interaction regime. 
   In this regime, polarization can drive notable relevant-irrelevant phase transition of the
sine-Gordon term \cite{Cui:2026}. 
  In the strong-coupling regime, by contrast, the gap grows large  and the characteristic of gapped spectral features become increasingly pronounced.  
  In particular, for small $q$ (indicated by the red dashed line in Fig.~\ref{Fig1}(b)), the system exhibits a linear gapless dispersion in the charge sector and a gapped dispersion in the spin sector, precisely the hallmark of a Luther–Emery liquid.
   As we shall show in Secs.\ref{Section III} and \ref{Section IV}, these distinct dispersion relations give rise to clearly separable peaks in the Bragg spectra, providing a direct experimental route to detecting this long-sought phase.

\section{Charge and Spin DSFs}\label{Section III}
To quantitatively connect the theoretical predictions with experimental observations, it is therefore essential to compute the DSF,  which encodes the spectral weight of low-energy excitations and forms the theoretical basis for interpreting Bragg spectroscopy measurements.  
Although the model Eq.\eqref{Hamiltonian} has been solved exactly, the calculation of correlation functions remains a long-standing challenge. 
The TLL theory provides a way to  approximatively calculate the correlation function of the 1D system at low-energy scale \cite{Giamarchi:2004}. 
In this section, we calculate  DSF of the low-energy excitations in the fully paired phase for both spin and charge sectors.

The charge excitations in the fully paired phase are gapless and obey a linear dispersion $\omega_c(q)=v_cq$ at low momenta. This is precisely the hallmark of a Tomonaga–Luttinger liquid. For higher momenta, the dispersion Eq.\eqref{Charge_Dispersion} takes the form of free fermion. Thus, the charge DSF can be well approximated by the expression for a non-interacting ideal Fermi gas \cite{Brand:2006,Hulet:2018,Hulet:2022}
\begin{eqnarray}
	S_c(q,\omega)=\frac{\text{Im}\chi(q,\omega)}{\pi(1-e^{-\beta\hbar\omega})},
\end{eqnarray}
where 
\begin{eqnarray}
	\text{Im}\chi(q,\omega)=\frac{Nm^{\ast}}{2\hbar^2q k_F}\pi\left(n_{q_{-}}-n_{q_+}\right),
\end{eqnarray}
with 
\begin{eqnarray}
	q_{\pm}=\frac{\omega m^{\ast}}{\hbar q}\pm\frac{q}{2},\ n_{k}=\frac{1}{e^{\beta(\epsilon-\mu)}+1},\ \epsilon=\frac{h^2k^2}{2m^{\ast}}.
\end{eqnarray}
The interaction effects encoded in the renormalized effective mass $m^{\ast}$ and  replacing $k_F$ with  the effective Fermi momentum $k_c=m^{\ast}v_c/\hbar$.

The spin sector presents a far more intricate problem. Unlike the charge mode, the spin excitations are gapped, which fundamentally alters the picture of TLL, and the calculation of the correlation functions is no longer trivial.   The challenge lies in the fact that generically the refermionized form for the correlation functions involves highly non-local operators. A way forward is offered by a remarkable property of the Luther–Emery liquid: at the special point $K_s=1/2$, the Sine-Gordon Hamiltonian that describes the gapped spin sector can be mapped exactly onto a free fermion model with a backscattering term. This is the essence of the Luther–Emery solution \cite{Luther_Emery:1974,Giamarchi:2004}, which we now exploit to obtain an analytical expression for the spin DSF. 

The starting point is the bosonized Hamiltonian for the spin sector
\begin{eqnarray}\label{Sine_Gordon}
	H_s &=&\frac{1}{2\pi}\int v_s\left[ K_s (\partial_x \theta_s(x))^2 + \frac{(\partial_x \phi_s(x))^2}{K_s} \right]dx\n
	&+& g\int  \cos(\sqrt{8}\phi_s(x))dx,
\end{eqnarray}
which is the familiar Sine-Gordon form, where $K_s,v_s$ are the Luttinger parameter and the velocity, respectively; $\phi_s, \theta_s$ are the bosonic field operator, and $g$ represents the interaction strength.
At the Luther-Emery line, i.e., $K_s=1/2$, this interacting bosonic theory is equivalent to a free fermion model with a backscattering term\cite{Giamarchi:2004}
\begin{eqnarray}
	H_s &=& \sum_k v_s k (c_{R,k}^\dagger c_{R,k} - c_{L,k}^\dagger c_{L,k})\n
	&+& \Delta  \sum_k c_{R,k}^\dagger c_{L,k} + \text{h.c.}, 
\end{eqnarray}
where $c_{R(L),k}$ is the annihilation  operator of the right-(left-)moving fermion, and  the energy gap $\Delta=\pi\alpha g$ is induced by the backscattering, with $\alpha$ a cut-off parameter in the TLL theory. While the gap $\Delta$ can be determined by the thermodynamical Bethe ansatz equations  (\ref{TBA_eq}).
This quadratic Hamiltonian can be diagonalized via a unitary  transformation that mixes right- and left-moving fermions
\begin{eqnarray}\label{U_D}
	\begin{split}
		\left(\begin{matrix}
			U_k\\
			D_k\\
		\end{matrix}
		\right)&=
		\left[\begin{matrix}
			\cos(\zeta_k) & \sin(\zeta_k)\\
			-\sin(\zeta_k) & \cos(\zeta_k)\\
		\end{matrix}\right]
		\left(	\begin{matrix}
			c_{R,k}\\
			c_{L,k}\\
		\end{matrix}\right),\\
		\zeta_k&=\frac{1}{2}\arctan2(\Delta,v_sk),\\
		\cos 2\zeta_k& =\frac{v_s k}{\sqrt{(v_s k)^2 +\Delta^2} },\quad \sin 2\zeta_k =\frac{\Delta}{\sqrt{(v_s k)^2 +\Delta^2} }.
	\end{split}
\end{eqnarray}
Then, the Hamiltonian can be diagonalized as
\begin{eqnarray}\label{H_two_ban}
	\begin{split}
		H_s &= \sum_{k}E_k\left(U_k^{\dagger}U_k-D_k^{\dagger}D_k\right),\\
		E_k&=\sqrt{\Delta^2+\left(v_sk\right)^2},
	\end{split}
\end{eqnarray}
which describes two independent bands separated by a gap $2\Delta$. If  the model is away from the Luther-Emery line, i.e., $K_s\ne1/2$, the interaction term gives a renormalization of the energy gap but the Hamilton still keeps the form of Eq.\eqref{H_two_ban}. The actual gap can be calculated from  the dressed energy equations (\ref{TBA_eq})  and  Bethe Ansatz equations (\ref{BA_eq}).  Physically, the operators $U^{\dagger}_k$ and $D^{\dagger}_k$ create excitations in the upper and lower bands, respectively. For the ground state $\Ket{0}$, the upper band is empty and the lower band is fully occupied, i.e.,
\begin{eqnarray}
	&\quad& \ev{U^{\dagger}_kU_{k^{\prime}}}_0=\ev{D_kD^{\dagger}_{k^{\prime}}}_0=0,\n
	&\quad& \ev{U_{k}U^{\dagger}_{k^{\prime}}}_0=\ev{D^{\dagger}_kD_{k^{\prime}}}_0=\delta_{k,k^{\prime}}.
\end{eqnarray}

With this diagonalized theory at hand, the calculation of the spin DSF becomes a well-defined problem in free fermion physics. In the k-space, the density operator can be expressed as
\begin{eqnarray}\label{rho_q}
	\rho(q)&=&\sum_{k}c^{\dagger}_{R,k+q}c_{L,k}+c^{\dagger}_{L,k+q}c_{L,k}\n
	&=&\sum_{k}\left(U_{k+q}^{\dagger}U_{k}+D^{\dagger}_{k+q}D_k\right)\cos\left(\zeta_{k+q}-\zeta_k\right)\n
	&\quad& +\left(U_{k+q}^{\dagger}D_k-D_{k+q}^{\dagger}U_k\right)\sin\left(\zeta_{k+q}-\zeta_k\right).
\end{eqnarray}
The time ordered density correlation function is defined as
\begin{eqnarray}\label{density_correlation}
	\chi(q,\tau)\equiv-\ev{T_{\tau}\rho(q,\tau)\rho(-q,0)}_0,
\end{eqnarray}
where $T_\tau$ is the time order operator, and $\rho(q,\tau)=e^{\tau H}\rho(q)e^{-\tau H}$ denotes the time evolution of the density operator in imaginary time. Substituting the Eq.\eqref{U_D} and Eq.\eqref{rho_q} into Eq.\eqref{density_correlation}, and after a straightforward but tedious calculation, the correlation function $\chi(q,\tau)$ is calculated as
\begin{eqnarray}
	\chi(q,\tau)=-\sum_ke^{-|\tau|\left(E_{k+q}+E_k\right)}\sin^{2}\left(\zeta_{k+q}-\zeta_k\right).
\end{eqnarray}
After a Fourier transform, we can get 
\begin{eqnarray}
	\chi(q,i\Omega)&=&\int d\tau e^{i\Omega\tau}\chi(q,\tau)\n
	&=&-\sum_k\sin^{2}\left(\zeta_{k+q}-\zeta_k\right)\frac{2\left(E_{k+q}+E_k\right)}{\Omega^2+\left(E_{k+q}+E_k\right)^2}.\n
\end{eqnarray}
The retarded correlation function is then obtained via the standard analytic continuation 
\begin{eqnarray}
	\chi^{\mathbb{R}}(q,\omega)=\chi(q,i\Omega\rightarrow\omega+i0^{+}),
\end{eqnarray}
and the DSF is given by the imaginary part of the retarded correlation function
\begin{eqnarray}\label{spin_dsf}
	&\ &S_s(q,\omega)=-\frac{1}{\pi}\text{Im}\chi^{\mathbb{R}}(q,\omega)\n
	&=&\sum_k\sin^{2}\left(\zeta_{k+\frac{q}{2}}-\zeta_{\zeta_{k-\frac{q}{2}}}\right)\delta\left(\omega-E_{k+\frac{q}{2}}-E_{k-\frac{q}{2}}\right).\n
\end{eqnarray}
This expression reveals the essential physics of the gapped spin excitations. Several important features follow immediately. First, excitations in this gapped system can only occur at energies satisfying
\begin{eqnarray}
	\omega&=&E_{k+\frac{q}{2}}+E_{k-\frac{q}{2}}\n
	&=&\sqrt{\Delta^2+u^2\left(k+\frac{q}{2}\right)^2}+\sqrt{\Delta^2+u^2\left(k-\frac{q}{2}\right)^2},\n
\end{eqnarray}
and at $k=0$, we can get the threshold 
\begin{eqnarray}\label{Threshold}
	\omega_{\text{th}}=\sqrt{4\Delta^2+v_s^2q^2}.
\end{eqnarray}
\begin{figure}[h]
	\centering
	\includegraphics[scale=0.6]{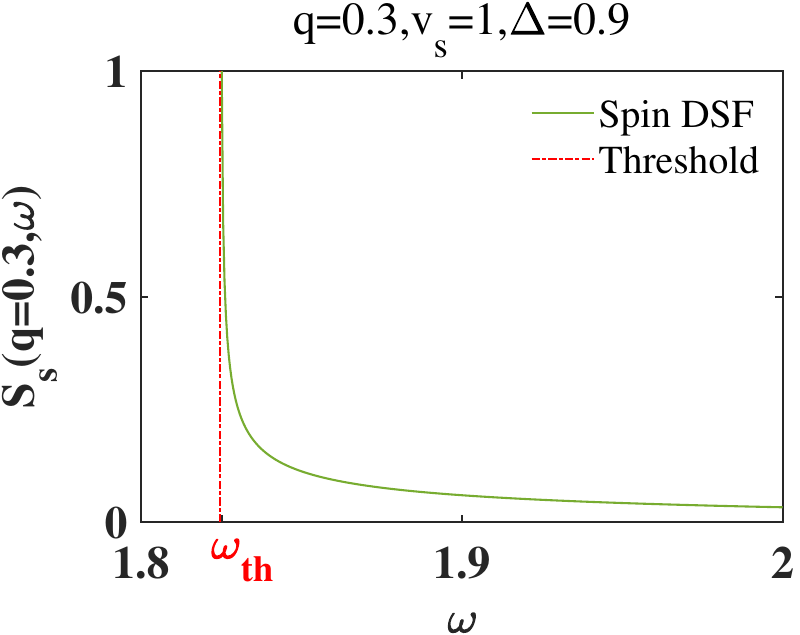}
	\caption{The green solid line shows the gapped spin DSF, which is determined by the Eq.\eqref{spin_dsf} with the parameter in  dimensionless unit $q=0.3,v_s=1$, and $\Delta=0.9$. The red dashed line shows the threshold of the excitation determined by the Eq.\eqref{Threshold}.} 
	\label{Fig2}
\end{figure}
In Fig.(\ref{Fig2}), we plot the  normalized spin DSF $S_s(q,\omega)$ given by Eq.\eqref{spin_dsf} as a function of $\omega$ for a fixed momentum transfer $q$.   A prominent feature immediately stands out: below the threshold frequency $\omega_{\text{th}}$(indicated by the red dashed line), the spin DSF vanishes identically. This complete absence of spectral weight at low energies is a direct and unambiguous consequence of the spin gap $\Delta$.  Physically, the gap implies that a minimum energy $\omega_{\text{th}}$ is required to create a spin excitation - any probe with energy below this threshold cannot excite the spin sector, and the system remains inert in its spin response.

This behavior stands in stark contrast to the charge sector, where the excitations are gapless and the DSF exhibits a finite response down to arbitrarily low frequencies. The dichotomy between the two sectors-a gapped spin response with a sharp threshold versus a gapless charge response with spectral weight extending to zero energy - constitutes the hallmark of the Luther-Emery liquid. It is precisely this contrasting behavior that makes Bragg spectroscopy a uniquely powerful tool for detecting the Luther–Emery phase. In the following section, we shall simulate the Bragg spectra under realistic experimental conditions and demonstrate how these distinct features manifest in the measured spin and charge dynamical response functions.

   	\section{Bragg spectra of the spin and charge mode}\label{Section IV}
   	 In this section, we simulate the Bragg spectra of the low-energy spin and charge excitations using experimental parameters adapted from Refs.\cite{,Hulet:2022,Hulet:2018,Hulet:2025}, and demonstrate how the characteristic signatures of the Luther-Emery liquid manifest in the Bragg response.  Firstly, we introduce the method for simulating Bragg spectra.
   	 
    In the experiment, fermions are prepared in a 2D optical lattice, which creates an array of quasi-1D tubes with a harmonic potential along the axial(x) direction. 
    We employ the LDA to treat the inhomogeneous trapped system. 
    The total particle number in the 1D tube is given by 
   	\begin{equation}
   		N(\mu_0)=\int n_{1D}(\mu,x)dx,
   	\end{equation}
   where $N(\mu_0)$ is the total particle number and $n_{1D}(x)$ is the density profile of the particle along the tube. 
   Solving the Eqs.~\eqref{TBA_eq} and ~\eqref{n_b_b_u}, we can numerically calculate $n_{1D}(x)$ for a given particle number $N$ and polarization $p$. 
   	
   	\begin{figure}[h]
   		\centering
   		\includegraphics[scale=0.33]{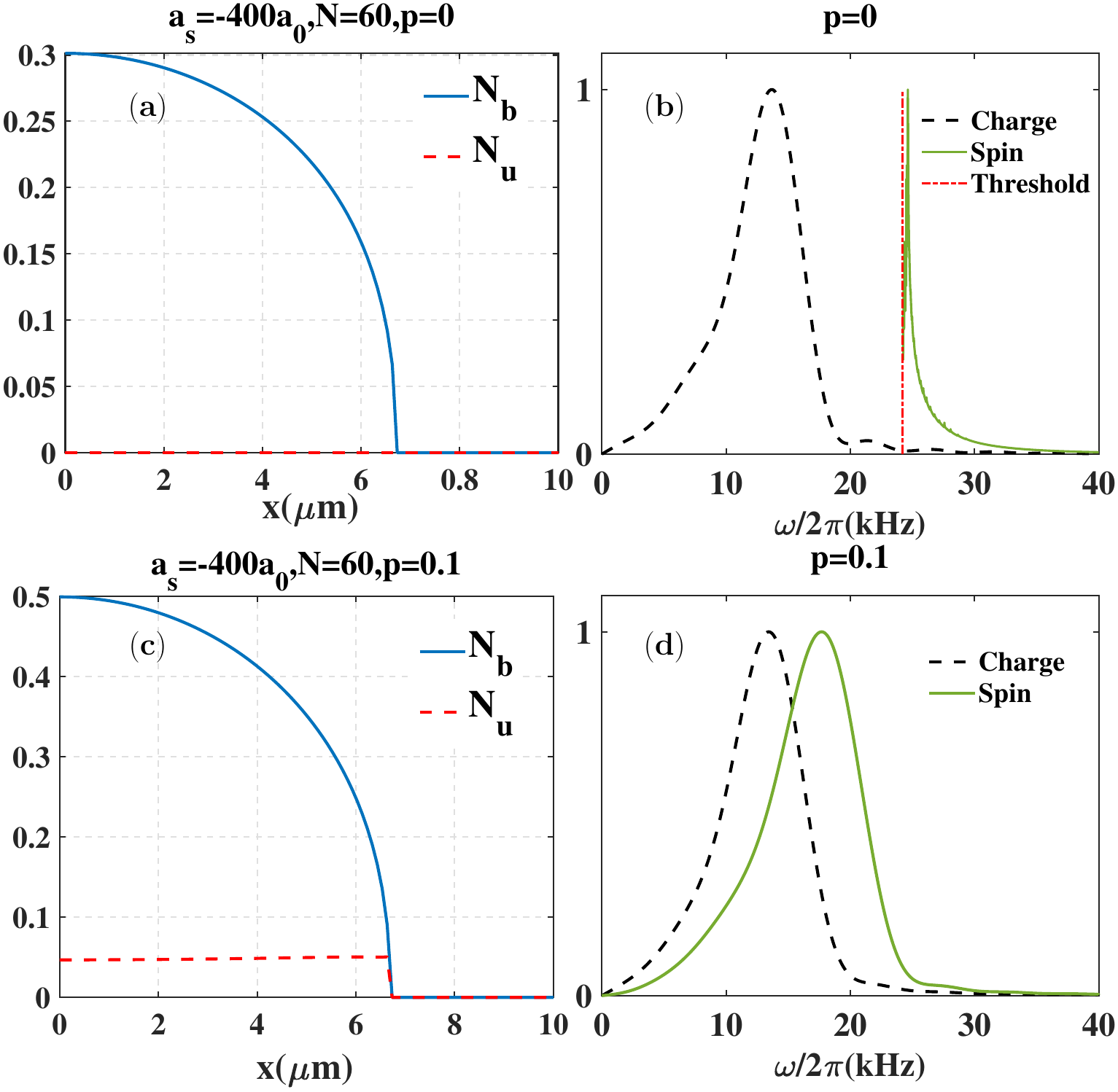}
   		\caption{ Bragg spectra and density profiles of a 1D two-component attractive Fermi gas in a harmonic tube.
The trap parameters are adopted from Ref \cite{Hulet:2022,Hulet:2025}, i.e., the  transverse and axial confinement frequencies are $\omega_{\perp}/2\pi\simeq227.5kHz$ and $\omega_x/2\pi=1.34kHz$, respectively.  At the low energy level, we set the temperature $T=0$, the total particle number $N=60$; and the length of the tube $L=20\mu m$. The 3D scattering length $a_s=400a_0$, which is related to the interaction strength via $c=-2\hbar^2/ma_{1D}$, with  $a_{1D}=(-a^2_{\perp}/2a_s)\left[1-1.4603(a_s/a_{\perp})\right]$ and $a_{\perp}=\sqrt{2\hbar/(m\omega_{\perp})}$ the transverse oscillator length. (a)  Density profile of the fermions in the tube with polarization p=0. The system is in the fully paired phase (blue solid line), while unpaired fermions (red dashed line) vanish. (b) Normalized Bragg spectra (DSF) of the spin (green solid line) and charge (black dashed line) modes in the fully paired phase with p=0 and  $q=1.47 \mu m^{-1}$. (c) Density profile of the  fermions in the tube with a small polarization $p=0.1$. The system enters the partially paired phase, where both paired (blue solid line) and unpaired (red dashed line) fermions coexist. (d) Normalized DSF of the spin (green solid line) and charge (black dashed line) modes in the partially paired phase with $p=0.1$  and $q=1.47 \mu m^{-1}$. }
   		\label{Fig3}
   	\end{figure}
   	
   	Furthermore,  the 1D average Bragg signal, which is essentially the DSF of  1D tube is calculated via
   	\begin{equation}
   	S_{1D}(q,\omega,N)=\int S(q,\omega,n_{1D}(x))n_{1D}(x)dx,
   	\end{equation}
   	where $S(q,\omega,n_{1D}(x))$ is the DSF at the local position.
   	The total DSF of a 3D sample is given by 
   	\begin{equation}
   		S_{3D}(q,\omega,N)=\sum_{i}S_{1D}(q,\omega,N(r_i)) N_{tube}(r_i),
   	\end{equation}
   	 where $N(r_i)$ is the average particle number of each 1D tube at the radial distance $r_i$ from the center of the sample, $i$ is the radial lattice index and $N_{tube}(r_i)$ is the number of tubes at the distance $r_i$. 
	 %

   	 Following the procedure outlined above, we calculate the Bragg spectra of  the 1D trapped attractive Fermi gas. We consider a harmonic tube containing 
   	 $N=60$ fermions with zero polarization $p=0$. 
	 Firstly, we determine the density profile of the trapped gas, as shown in Fig.~\ref{Fig3}(a), where only paired fermions are
   	  present - confirming that the system resides in the fully paired phase. 
	  The resulting spin and charge Bragg spectra are presented in Fig.~\ref{Fig3}(b), denoted by the green solid and black dashed lines, respectively. 
	  A salient feature immediately stands out: the spin Bragg peak is shifted to a significantly higher frequency than the charge peak, and the spin gap manifests as a clear and measurable suppression of spectral weight below the threshold (indicated by the res dashed line), in stark contrast to the gapless charge response that extends down to zero energy. This is a direct consequence of the spin gap $\Delta$ as being discussed  in Sec.\ref{Section III}.
	  The clear difference of the spectral structure  between the two modes thus provides an unambiguous experimental signature that distinguishes the Luther-Emery liquid from both the conventional Luttinger liquid (where both sectors are gapless) and other gapped phases.
	  Furthermore, as demonstrated in Fig.~\ref{Fig1} (c) and (d), both the energy gap $\Delta$ and the spin velocity $v_s$ increase monotonically with the coupling strength $|\gamma|$. Consequently, the threshold frequency $\omega_{\text{th}}=\sqrt{4\Delta^2+v_s^2q^2}$ grows with increasing interaction, pushing the onset of spin excitations to higher energies and making the spectral separation between spin and charge modes more pronounced. This tunability is of particular experimental relevance: by adjusting the interaction strength via Feshbach resonances, one can optimize the experimental conditions to maximize the visibility of the Luther–Emery signature.
   	  
   It is crucial to stress that experimental observation of the aforementioned signature poses substantial technical challenges.
Realizing a fully paired, perfectly unpolarized many-body system constitutes an enormous experimental hurdle: residual impurities, thermal fluctuations, and small  imbalances in spin populations alike can generate unpaired fermionic species.
Even a tiny finite polarization $p>0$ pushes the system into a partially paired phase, within which both spin and charge excitations turn gapless \cite{J_F_Pan:2023}.
In this regime, the spin gap vanishes. Without the characteristic frequency splitting that separates the two Bragg peaks, the hallmark Luther–Emery signature is obscured \cite{Hulet:2025}.

To illustrate this argument and enable direct, unambiguous comparison, we plot the Bragg spectra for a weakly polarized system with 
$p=0.1$ in Fig.~\ref{Fig3}(d) using the method in ~\cite{Hulet:2025}.
The associated density profile is displayed in Fig.~\ref{Fig3}(c), which captures the coexistence of paired and unpaired fermions.
Within this partially paired phase, the spin and charge Bragg peak frequencies lie close to each other-markedly distinct from the widely split peaks characteristic of the fully paired state.
This side-by-side comparison demonstrates that the frequency splitting between spin and charge Bragg resonances acts as a sensitive probe for the LEL: a large spectral separation unambiguously signals a gapped spin sector, whereas closely spaced peaks correspond to the gapless partially paired phase.
Accordingly, unambiguous experimental observation of the LEL requires stringent suppression of spin polarization to stabilize the system within the fully paired regime.
   	  
  \section{Conclusion}\label{Section V}
  
  In this work, we have analytically investigated the low-energy excitations of a one-dimensional attractive Fermi gas in the fully paired phase and proposed an experimental protocol for identifying the Luther-Emery liquid via Bragg spectroscopy. By solving the Yang-Gaudin model with the Bethe ansatz, we have obtained exact results for the gapless charge and gapped spin excitation spectra. Within the bosonization framework, we have derived low-energy approximate expressions for the DSF of both spin and charge sectors, which permits quantitative numerical simulations of Bragg spectra. Our results  reveal several key signatures of the LEL. First, the spin Bragg peak is shifted to higher frequencies by the spin gap $\Delta$, whereas the charge peak remains at lower frequencies - this distinct spectral splitting constitutes a hallmark of the Luther-Emery phase. Second, the spin gap manifests as a sharp threshold in the spin DSF, below which excitations are completely forbidden, in stark contrast to the gapless charge response that extends down to zero energy. Third, the magnitude of the frequency separation grows with interaction strength, making the LEL signature increasingly pronounced and experimentally accessible in the strong-coupling regime. Crucially, we have shown that even a small polarization $p>0$ drives the system into the partially paired phase, where the spin gap closes and the characteristic spectral splitting between spin and charge excitations is minimal, indicating  a charge-charge separation, thereby masking the LEL signature. 
  Thus we conclude that in order to realize the LEL, it is important to eliminate spin imbalance.
  Taken together, our findings establish a rigorous theoretical framework alongside a practical experimental pathway toward  a observation of the Luther–Emery liquid in ultracold atomic quantum gases.

	\bibliographystyle{apsrev4-2}

\end{document}